\documentclass[sigconf,anonymous=False]{acmart}
\renewcommand\footnotetextcopyrightpermission[1]{} 
\AtBeginDocument{%
  \providecommand\BibTeX{{%
    \normalfont B\kern-0.5em{\scshape i\kern-0.25em b}\kern-0.8em\TeX}}}

\acmConference[archix ’22]{}{ July 11–15,
  2022}{}
\usepackage{array}
\usepackage{caption}
\usepackage{subcaption}
\usepackage{amsfonts}
\DeclareUnicodeCharacter{2212}{-}
\begin{document}

\title{SML: Enhance the Network Smoothness with Skip Meta Logit for CTR Prediction}

\author{Wenlong Deng}
\email{dengwenlong.dwl@bytedance.com}
\affiliation{%
  \institution{TikTok}
  \country{China}
}
\author{Lang Lang}
\email{langlang.42@bytedance.com}
\affiliation{%
  \institution{TikTok}
  \country{China}
}
\author{Zhen Liu}
\email{liuzhen.raullz@bytedance.com}
\affiliation{%
  \institution{TikTok}
  \country{China}
}
\author{Bin Liu}
\email{bin.liu@bytedance.com}
\affiliation{%
  \institution{TikTok}
  \country{US}
}


\begin{abstract}
Click-Through Rate (CTR) prediction plays a core role in the recommendation and online advertising system. 
Theoretically, deep neural networks$\footnote{DNN here refers to fully-connected feedforward network(FFN). This is also called  multi-layer perceptron network or dense network in some other literature. Since in CTR prediction there is a higher order cross layer network, unless otherwise specified, DNN in our paper refers to the FFN network architecture.}$ (DNN) embrace the ability to express complex functions thereby widely applied in ctr prediction models.  
However, there is no guarantee that DNN could converge to the expected function.  
Extensive research in CTR has focused on designing brand new cross layers as an enhancement while incorporating DNN in an ensemble manner.
Although DNN is still the fundamental building block for almost all ctr models, research on optimization of its convergence has received far less attention.
\\
In light of the smoothness property brought by skip connections in ResNet, this paper proposed the Skip Logit to introduce
the skip-connection mechanism that fits arbitrary DNNs' dimensions and embraces similar properties to ResNet. Meta Tanh Normalization (MTN) is designed to learn variance information and stabilize the training process.  With these delicate designs, our Skip Meta Logit (SML) brought incremental boosts to the performance of extensive SOTA ctr prediction models on two real-world datasets. In the meantime, we prove that the optimization landscape of arbitrarily deep skip logit networks has no spurious local optima. 
Finally, SML can be easily added to building blocks and has delivered offline accuracy and online business metrics gains on
app ads learning to rank systems at xxxx.

\end{abstract}



\maketitle
\section{Introduction}

The accuracy of click-through rate (CTR) prediction has remained to be one of the most
important problems in search, recommendation systems\cite{Rec_book} and computational advertising\cite{cadv}. Due to the general success of deep neural networks(DNNs) in different domains\cite{DNN}, DNNs have been extensively studied and deployed in real production environments for CTR prediction. 

A vast majority of deep CTR models stem from Wide\&Deep Learning (WDL)\cite{wide_deep}, which firstly state complementarity of memorization of wide linear models and generalization of DNNs, and integrate both parts to improve model expressiveness.
Although DNNs are considered to be universal function approximators\cite{appx} and their expressive power grows exponentially with depth\cite{expressive}, there is no strict guarantee of their convergence\cite{fail_dnn_gd}.
In addition, a large number of studies have proved that a pure DNN is insufficient in CTR prediction\cite{pnn, deepfm, dcn, dcnv2, xdeepfm}.
Following the ideas of WDL, massive efforts have been put into designing new architectures to patch on DNN.
For example, different vector operations are used to enrich calculation of model weights\cite{pnn}; multiple variants of matrix factorization mechanisms are used to express second-order cross information\cite{deepfm,afm,nfm}; some well-designed structures such as CIN\cite{xdeepfm} and CrossNet\cite{dcn, dcnv2} can express polynomial computations; self-attention mechanisms\cite{auto_int} are used for feature fusion.
Various ideas lead to considerable progress. 
However, optimization of DNN itself has been consistently neglected to some extent in the domain of ctr prediction.
This basic building block's potentiality still awaits to be aroused. 

How to enhance the smoothness of DNN to improve model smoothness? 
As a fundamental question, there is a rich history of prior work addressing this problem from different perspectives such as model architecture, weight distribution and optimization algorithm\cite{resnet_v1,batchnorm,adam}.
Among these, skip connection mechanism introduced by ResNet\cite{resnet_v1} is one of the most popular methods for its simplicity yet effectiveness.
This mechanism is proven to embrace the abilities to enhance the smoothness of network\cite{landscape,resnet_convex} and prevent the rank collapse\cite{attention_skip}. 
These properties are beneficial to the stability of the propagation process and make it possible to train deeper networks, thereby improving the model's expressive ability.
As a result, Resnet has been widely used in the field of computer vision\cite{resnet_v1} and natural language processing\cite{attention,attention_skip}.
However, there are two obstacles to directly applying ResNet to a CTR prediction model.
First of all, the original form of ResNet strictly requires the number of hidden units in each layer to remain identical\cite{resnet_v2}, which cannot accommodate the tower-like DNNs that is a very common setup in ctr models. 
Secondly, Trained networks are far more sensitive to their lower layer weights which are not robust to noise\cite{expressive} and Batch Normalization(BN) is a necessary technique in ResNet to stabilize network variance \cite{Soham,resnet_v2}. 
It uses the distribution of the current minibatch as a proxy for the distribution of the entire dataset\cite{online_bn}. 
However, in industrial settings, online learning is often the default setting to achieve real-time performance \cite{online}. 
This means that the batch size is either shrunk to a very small even one, or indeterminate. 
This makes it impossible to obtain stable statistics for a batch of data to apply BN.

In this paper, We flatten the ResNet into Skip Logit to fit dnn of arbitrary shape and our theoretical analysis reveals that Skip Logit networks are free of spurious local optima. As a result, the first obstacle is solved without losing smoothness property.
We further prove that Skip Logit won't suffer from the network degradation issue since the output variance increases linearly with the depth. 
Regarding the problem that trained networks are far more sensitive to their lower layer weights\cite{expressive}, Meta Tanh is proposed to be well compatible with online learning while stabilizing the training process by adaptively rescaling the variance of each layer, thus solving the second obstacle above.
Combining with Skip Logit and Meta Tanh, our final proposed SML network can be trained stably at a depth of more than 100 layers, while the DNN collapses in less than 30 layers. 
In addition to good convergence and stability, comprehensive experiments on two real-world datasets indicate that SML is a board-spectrum approach to improve DNN,  which generally pushes SOTA deep ctr models one step further on performance.

We summarize our main contributions as follows:
\begin{itemize}
\item We propose SML, a novel skip-connection mechanism to refine convergence of DNN. To our best knowledge, this is the first work to systematically investigate how to improve the smoothness of DNNs in the area of ctr prediction. 
\item We establish a theoretical explanation of SML from the perspective of optimization landscape in detail. Skip Logit is proved to have no spurious local optima under specific conditions; Meta Tanh was shown to effectively control the variance of the model to be linear to the depth.  Ablation studies demonstrate the improvement brought by each design.
\item We conduct extensive experiments on two real-world datasets and show that SML can ubiquitously improve the performance of SOTA deep CTR models. In addition, property analysis provides empirical progress on how DNN works for ctr prediction.
\end{itemize}


\section{Background}
In this section, we first introduce related works from three fields, residual network, normalization and meta learning respectively. Then we give a quick review of ResNet and its variance analysis.
\subsection{Related Work}
{\bf Residual Network}(ResNet) is a milestone in deep learning. ResNet is equipped with skip connections between layers and exhibits the forward and backward signals can be directly propagated from one block to any other block\cite{resnet_v2}. Recent study on investigating the optimization impact of ResNet behind is well established. Hardt \textit{et al.}\cite{imidl} proofed arbitrarily deep linear residual networks have no spurious local optima. Liu \textit{et al.}\cite{TUIS} proofed under certain assumptions, residual combined with proper normalization avoids being trapped by the spurious local optimum and converges to a global optimum in polynomial time. Furtherly, Soham etal\cite{Soham} show that normalization downscales the residual branch relative to the skip connection, which ensures that, early in training, the function computed by normalized residual blocks in deep networks is close to the identity function. Some other papers also state skip connection can increase generalization ability. Andreas\textit{et al.} \cite{ensemble} declare ResNet behave like ensemble of relatively shallow network. He \textit{et al.}\cite{Fengxiang} demonstrate that residual connections may not increase the hypothesis complexity of the neural network compared with the chain-like counterpart and give a generalization bound for ResNet. However, the use of residual connection is still under-researched in the recommendation system.
\\
{\bf Normalization} techniques are of vital importance to stably and effectively train deep neural networks (DNNs) \cite{batchnorm,ba2016layer,Wu_2018_ECCV}. Arguably, the most popular of these is BatchNorm\cite{batchnorm}, whose success can be attributed to several beneficial properties that allow it to stabilize a DNN’s training dynamics. Recently studies also linked normalization with ResNet, \cite{beyondbatch,Soham} illustrates activations-based normalization layers can prevent exponential growth of activations in ResNets. However, for batch-level normalization mechanisms, the calculation of mean and variance relies on the whole mini-batch. The effectiveness of normalization may degrade when the batch size is not sufficient to support the statistics calculation. To ease the issue, more recently, Layer Normalization (LN)\cite{layer_norm} operates along the channel dimension and standardizes the features from a single mini-batch by its mean and variance. It can be used with batch size 1. Instance Normalization (IN)\cite{Instance_norm} standardizes each feature map with respect to each sample. However, Ekdeep \textit{et al.}\cite{beyondbatch} found small group sizes result in large gradient norm in earlier layers, hence explaining training instability issues in Instance Normalization.
\\
{\bf Meta Learning} has been used in normalization to fit the instance-level normalization and address a learning-to-normalize problem. Jia etal \cite{Jia2019} proposed instance-level meta (ILM) normalization to associate the rescaling parameters with the input feature maps rather than deciding the parameters merely from back-propagation. Fan etal \cite{Fan_2021_CVPR} introduced the learning of standardization statistics to complement the missing part in \cite{Jia2019} which can be viewed as a generic form of the traditional normalizations. Park etal \cite{metavariance} proposed Meta-Variance Transfer (MVT) which learns to transfer factors of variations from one class to another meta variance and improves the overall classification performance. Observing the success of meta learning, this paper will focus on the variance in standardization and investigate its benefit in the field of CTR training. 

\subsection{Preliminaries} 
The ResNets developed in \cite{resnet_v1} are modularized architectures that stack building blocks of the same connecting shape. for any deeper unit $L$ and any shallower unit $l$ , the output can be represent as:
\begin{equation}
x_{L} = x_{l} + \sum^{L-1}_{i=l} F(x_{i},W_{i})
\end{equation}  
Without normalization on the residual path, the variance of residual path is:
\begin{equation}
Var(x_{l+1}) = E[Var(F(x_{l})|x_{l})] + Var(x_{l})
\end{equation}  
regarding the decay of variance after Relu:
\begin{equation}
Var(Relu(x)) \leq \delta^{2}(1-\frac{2}{\pi})
\end{equation}  
(proof in \ref{variance}), we combined with initialization methods such as He \textit{et al.}\cite{ini_he}, the output variance of each residual branch will be about the same as input variance, and thus:
\begin{equation}
Var(x_{l+1})\approx 2Var(x_{l})
\end{equation}  
 As a result, the variance of each layer will be $2^{l}$. This exponential variance growth rate with depth will cause explosion in training.


\begin{figure}
    \centering
    \includegraphics[width=.9\linewidth]{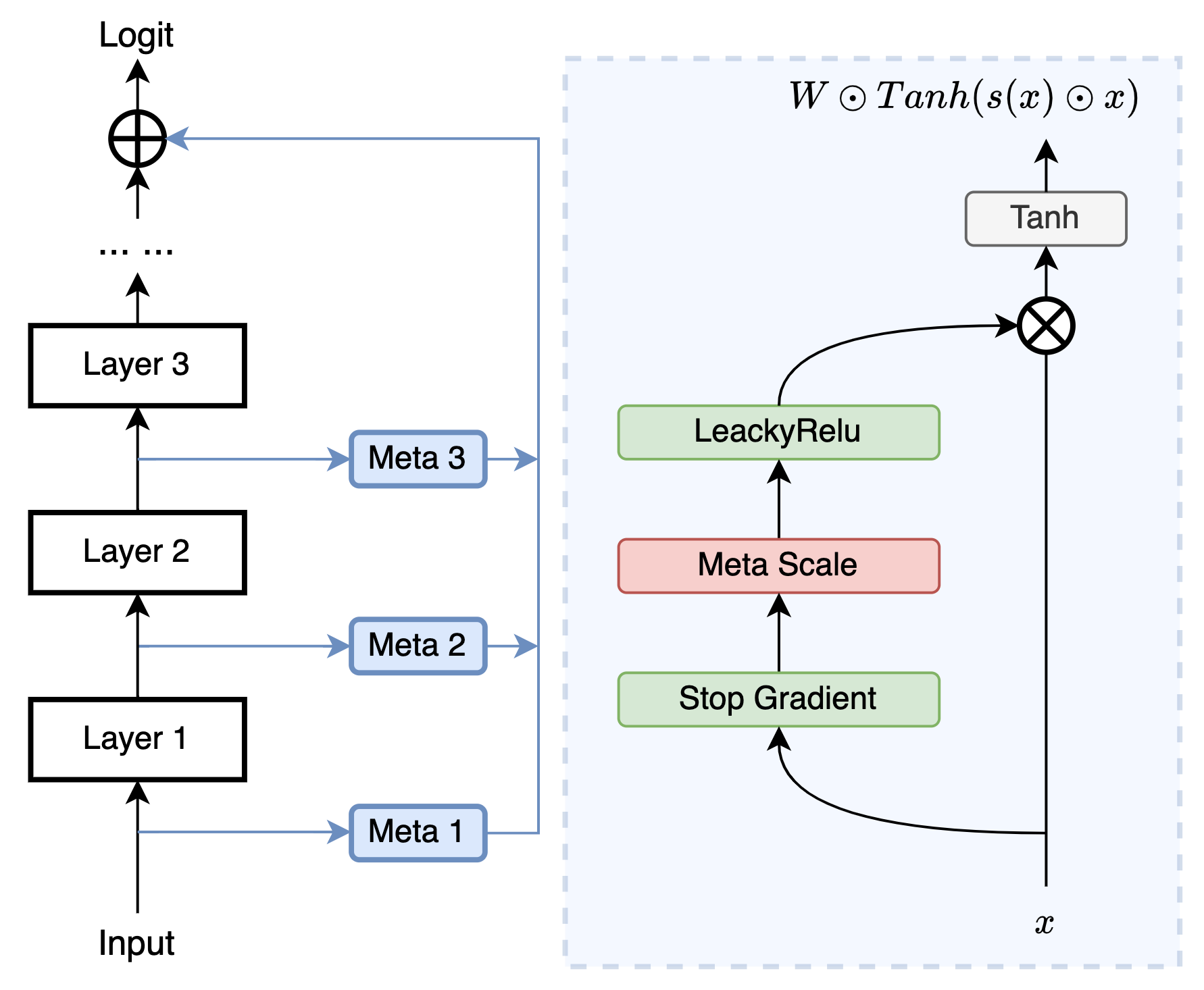}
    \caption{The framework of Skip Meta Logit(SML)}
    \label{fig:SML}
\end{figure}

\section{Methodology}

In this section, we detail two major improvements, Skip Logit and Meta Tanh Normalization respectively. 
We firstly set a theoretical footing that variance of Skip Logit's will be a linear growth rate with depth, and then give optimization landscape of Skip Logit schema under certain assumptions.
Finally, we study Meta Tanh normalization and demonstrate its benefits on training in terms of stability, optimization and variance.

\subsection{Flat Resnet to Skip Logit}
ResNet\cite{resnet_v1} structure prevents layer outputs and gradients from vanishing by forcing the variance to grow with depth\cite{fixup}. 
The original format of ResNet directly sums up the output of the residual path and the main path. 
As illustrated in the preliminaries, normalization should be added to residual path to avoid an exponential increasing of variance. 
Considering batch normalization cannot obtain the statistical information in online serving and the training instability issues in non-batch normalization methods\cite{beyondbatch}, we flat the skip connection and directly add the output of each skip path into the final output, i.e, the logit value in ctr prediction problem. 
We named it Skip Logit.
The formula is given by:
\begin{equation}\label{formu}
logit = x_{0}+ \sum_{i=1}^{L-1}F(x_{i},W_{i}^{L-1})
\end{equation}  
Where $x_i$ is the output of the i-th layer, $W_i$ is the trainable weights involved in the i-th skip path, and $L$ is the total number of skip paths. 
At first glance, Skip Logit can bring two distinct advantages.
First, this design is applicable to DNN of arbitrary shape.
Second, the Skip Logit allows shallow layers directly influence the final output, which in turn the gradients can flow back to each shallow layer in backpropagation, thus emphasizing the status of shallow layers thus benefits the optimization of model\cite{fixup,Soham,TUIS}.

\subsubsection{Linear Variance Growth}
Combining equation \ref{formu} and He\cite{ini_he} initialization, the growth of variance at the output layer is then stereo which is a linear rate:
\begin{equation}\label{nvar}
Var(f(x)) = L * Var(x)
\end{equation}    
Linear variance growth rate helps avoiding the explosion of gradient.
Recall that the variance growth rate is exponential to the the depth in original ResNet, which constitutes the primary limitation of the network depth.

\subsubsection{Optimization Landscape of Linear Skip Logit}
Under linear condition, we prove the proposed networks can have no \textit{critical point} other than the global optimum. 
Before proving the theorem, we give some notation and borrow one key lemma from \cite{imidl}:
\\
Consider the problem of learning a linear transformation $R: R^{d} \rightarrow R^{d}$ from noisy measurements $y=Rx+\xi$, where $\xi \in \mathcal{N} (0,I_{d})$ is a $d-$dimensional spherical Gaussian vector. Denoting by $D$ the distribution of the input data $x$, let $\Sigma= E_{x\backsim D}[xx^{T}]$ be its covariance matrix. Our goal is to gain insights into the optimization landscape of neural nets, and in particular, skip logit networks.
Recall that $||A_{i}||$ is the spectral norm of $A_{i}$. We define the norm $|||.|||$ for the tensor $A$ as the maximum of the spectral norms of its slices,
\begin{equation}\label{t21}
|||A||| := max_{1\leq i \leq l}||A_{i}||
\end{equation} 
The lemma of this section states that the objective function $f$ has an optimal solution with small $|||.|||$-norm. 
\\
{\bf Lemma 1:} \textit{For a ResNet $f(A)$ with $l$ shortcuts, Suppose $l/3 \geq \gamma$ and $det(R) > 0$. Then, there exists a global optimum solution $A^{\ast}$ of the population risk $f(.)$ with norm}
\begin{equation}\label{t21}
|||A^{\ast}||| \leq (4\pi + 3\gamma)/l
\end{equation} 
where \textit{$\gamma:=max\{|log\delta_{max}(R)|,|log\delta_{min}(R)|\}$}.
This lemma is from the proof of optimization landscape of linear residual networks in \cite{imidl}. 
Based on Lemma 1, the Thereom 1 is given:
\\
{\bf Theorem 1:}\label{theorem1} \textit{for any $\gamma_{i} < 1$, we have that any critical point A of the objective function $f$ must also be a global minimum.}
\begin{equation}\label{t21}
||\bigtriangledown f(A)||^{2}_{F} \geq 4\sum^{l}_{i=1}(1-\gamma_{i})^{2} \delta_{min}(\Sigma)||f(A) - C_{opt}||^{2}_{F}
\end{equation} 
\\
Where $||.||_{F}$ denote the Frobenius norm of a matrix, $\delta_{min}(\Sigma)$ denote the minimum singular value of .
For short, from Theorem 1 we know that if $A$ is a critical point, namely, $\bigtriangledown f(A) = 0$, then we have that $f(A) = C_{opt}$. 
That is to say, arbitrary critical point $A$ is a global minimum.
This result reveals that introducing Skip Logit to DNN will enhance the smoothness of loss landscape, thereby ease the training procedure.
The proof of Theorem 1 is detailed \ref{theorem1}. 

\subsection{Meta Tanh Normalization}
Skip Logit directly connect each layer to the final output has the nice property to enhance optimization of the network. 
As the point raised in this work\cite{expressive}, the shallow layers is critical to model performance while there weights suffers much less robust to noise, a proper normalization method is in need to stabilize the training process.
The observation of \ref{logit_dist} also agrees this point of view.
Therefore, the Meta Tanh Normalization(MTN) is designed to well train the shallow layers and alleviate the influence from noise. 
Equation \ref{ATN_for} gives the formula of MTN:
\begin{equation}\label{ATN_for}
x_{norm} = W\odot Tanh(s(x)\odot x)
\end{equation}
Where $W$ is the scale parameter and $s(x)$ is the meta variance learner. 
MTN introduces three main benefits:

\begin{table*}
\centering
\caption{Performance Comparision}
\begin{tabular}{l c c c c c c c c}
\toprule
               & \multicolumn{4}{c}{Criteo}                                           & \multicolumn{4}{c}{Avazu}  
\\ \cline{2-5} \cline{6-9} 
{\bf Methods}   & \multicolumn{2}{c}{w/o SML}                                           &                      \multicolumn{2}{c}{w/ SML}  & \multicolumn{2}{c}{w/o SML}                                           &                      \multicolumn{2}{c}{w/ SML}  
\\ 
                                 & logloss                            & AUC                        & logloss                            & AUC      
                                  & logloss                            & AUC      
                                   & logloss                            & AUC      
                                   \\ \midrule
DNN & 0.44218 & 0.80948 & 0.44123 & 0.81042 & 0.37779 & 0.78376 & 0.37727 & 0.78459           \\ 
DCN & 0.44224 & 0.80936 & 0.44144 & 0.81021 & 0.37776 & 0.78374 & 0.37738 & 0.78453           \\ 
DCN-V2 & 0.44029 & 0.81143 & 0.44013 & {\bf 0.81171} & 0.37697 & 0.78499 & 0.37654 & {\bf 0.78583}           \\ 
DeepFM & 0.44216 & 0.80949 & 0.44146 & 0.81026 & 0.37767 & 0.78388 & 0.37716 & 0.78482           \\ 
XDeepFM & 0.44129 & 0.81046 & 0.44099 & 0.81086 & 0.37753 & 0.78436 & 0.37674 & 0.78541           \\ 
AutoInt & 0.44218 & 0.80947 & 0.44143 & 0.81025 & 0.37764 & 0.78411 & 0.37749 & 0.78435           \\ \bottomrule
\end{tabular}
\\
\label{compare}
\end{table*}

\subsubsection{Stable Training}
Tanh function has the mathematical properties of being smoothly differentiable and mapping outliers toward the mean\cite{jurafsky2000speech}. 
Since the maximum absolute value is capped to 1, we gave the out variance with Tanh activation.
\begin{equation}\label{tvar}
Var(tanh(x)) \leq 1
\end{equation}
Compared with Relu activation, different skip paths will have a more stable gradient with respect to weights. 
What's more, when the input is an outlier, the main path will dominate the variance and the shallower skip path won't be sensitive to perturbations which increased the stability\cite{expressive}. 
With variance standardization, according to Equation \ref{nvar} and \ref{tvar}, the variance can be updated to a more stable growth of variance at the final output layer:
\begin{equation}
\label{mtn_var}
Var(f(x))  \leq  Var(x)+L-1 
\end{equation}
From the first term of equation \ref{mtn_var}we can see, the variance in the original input is well preserved. 
From the second term we can see that by decoupling the model depth from the variance of the input data, the amplification effect of the network on $Var(x)$ is further suppressed.

\subsubsection{Optimization}
Shortcut has been proofed being of vital importance for optimization in both our theorem 1 and \cite{resnet_v2}, which leads to nice backward propagation properties. 
From equation \ref{formu} and denoting the loss function as $\epsilon$, follow the chain rule of backpropagation:
\begin{equation}
\frac{\partial \epsilon}{\partial x_{l}} = \frac{\partial \epsilon}{\partial x_{L}} \frac{\partial x_{L}}{\partial x_{l}} = \frac{\partial \epsilon}{\partial x_{L}}(tanh + \frac{\partial}{\partial x_{l}}\sum^{L-1}_{i=l}F(x_{i},W_{i}))
\end{equation} 
The formula suggests that the signal can be directly propagated from any unit to another both forward and backward, with an extra $tanh$ regularization to avoid acute parameter change. We can further relax the formula with Taylor expansion of $tanh$:
\begin{equation}
tanh(x) = x + O(x^{2})
\end{equation} 
and got equation \ref{tal_tanh}:
\begin{equation}\label{tal_tanh}
\frac{\partial \epsilon}{\partial x_{l}} \approx \frac{\partial \epsilon}{\partial x_{L}}(1 + \frac{\partial}{\partial x_{l}}\sum^{L-1}_{i=l}F(x_{i},W_{i})),
x_{l} \ll 1
\end{equation} 
This indicates the addition of Tanh can be linear when input $x$ is small. With appropriate scaling, Tanh can be used to provide a linear transformation for input values in the neighborhood of “expected” values while reducing values that are outside the expected range\cite{tanh_transform}. As a result, our proposed network can achieve strong optimization and stable training. 
\subsubsection{Variance Scaling}
To instantly learn the input's variance and well scale the input to achieve a linear transformation, we use a meta learner $s(x)$\cite{maxout,Fan_2021_CVPR} to achieve the goal. 
To lower the computational cost and leverage the original statistics information contained in the feature maps, the standardization network chooses a stop gradient input, instead of the original feature map, as the input to learn the standardization statistics:
\begin{equation}\label{eq2}
s(x) = leakyrelu(w\times{x^{'}})
\end{equation}
Where $w$ is a linear layer and $x'$ is the stop gradient of input $x$ which eases the training procedure. Leaky Relu\cite{leakyrelu} introduces sparsity to Tanh output while reducing the number of dead kernels. The variance scaling will additionally alleviate the saturation. Experimental proof can be found in ablation study \ref{meta_analysis}.



\section{EXPERIMENT}
In this section we aim to answer the following research questions:
\begin{itemize}
\item {\textbf{RQ1}}: Can SML improve the performance of sota baseline models?
\item {\textbf{RQ2}}: What is the effect of each design in SML on the model performance, such as Skip Logit, activation function, and Meta Scale Method?
\item {\textbf{RQ3}}: What is the empirical reason that SML helps the model to converge and stabilize?
\item {\textbf{RQ4}}: How does the SML network perform in a real industrial environment? 
\end{itemize}

We first describe the experiment setup, including training
datasets, baseline approaches, and details of the hyper-parameter
search and training process, then answer the above questions with experimental results.

\subsection{Settings}
\subsubsection{Datasets}

We conduct experiments on two datasets.
{\bf Criteo}$\footnote{https://www.kaggle.com/c/criteo-display-ad-challenge}$ is a widely used click-through-rate (CTR) prediction benchmark dataset that contains user logs over a period of 7 days. It contains 45M samples with 13 continuous features and 26 categorical features. We
follow AutoInt\cite{auto_int} to preprocess data.
{\bf Avazu}$\footnote{https://www.kaggle.com/c/avazu-ctr-prediction/overview/description}$ contains 10 days of click logs. It has a total of 23 fields with categorical features including app id, app category, device id, etc. The total number of samples is 40M.

We explicitly follow the pre-processing procedure reported in AutoInt\cite{auto_int}. Then both datasets are split to 8:1:1 as the training set, validation set, and test set. The validation set is only used for tunning hyper-parameters, and all experiment results are reported based on the test set.

\subsubsection{Competing Models}
We examine the effects of SML on widely used deep ctr models including: 
\\
{\bf DeepFM} \cite{deepfm} is an extension of Wide\&Deep\cite{wide_deep} that substitutes LR with FM to explicitly model second-order feature interactions.
\\
{\bf XDeepFM}\cite{xdeepfm} proposes to capture high-order
feature interactions in a vector-wise way via a compressed interaction network (CIN).
\\
{\bf AutoInt}\cite{auto_int} The interaction layer of AutoInt adopted the multihead self-attention mechanism. For simplicity, we assume a single
head is used in AutoInt; multi-head case could be compared summarily using concatenated cross layers.
\\
{\bf DCN}\cite{dcn} In DCN, a cross network is proposed to perform highorder feature interactions in an explicit way. In addition, it also integrates a DNN network following the Wide\&Deep framework.
\\
{\bf DCN-V2}\cite{dcnv2} proposed across network to perform highorder feature interactions in an explicit way. In addition, it also integrates a DNN network following the Wide\&Deep framework.

\subsubsection{Training Settings}
All the baselines and our approaches are implemented in TensorFlow v2. For a fair comparison, all the implementations were identical across all the models except for the feature interaction component. We apply Adam with a learning rate of
0.001 and a mini-batch size of 10000. The default number of MLP neurons is set to 1024, 512, 256 for Criteo and 1024, 512 for Avazu. Batch normalization\cite{batchnorm} and drop-out\cite{dropout} are disabled in the network because the former cannot acquire accurate statistical information in an online serving environment and the latter lacks reproducibility and decreases the stability of the system in practice. All the other hyperparameters are tuned on the validation set. 
Once the model setting is settled, we run the experiments 5 times and report the average value.

\subsection{Performance Comparison(RQ1)}
We evaluate baseline models with and without applying SML on two popular datasets Criteo and Avazu, the results are given in table \ref{compare}. 

One primary observation is that SML universally improves performance regardless of the model architecture or dataset used. On one hand, optimization on DNNs can still bring considerable room for improvement.  On the other hand, the universality of SML implies that the benefits of SML and feature crossing architectures are compatible with each other. Our work may open up an orthogonal route for the optimization of ctr models. 

There is also another interesting observation that standalone DNN w/ SML is able to beat quite a few complicated baseline models. Specifically, DNN w/ SML is only weaker than DCN-V2 on the avazu dataset; DCN-V2 and xDeepFM on the criteo dataset. This supports our view that Meta Tanh has the side effect of introducing interactive information, which to some extent makes DNNs more expressive. We dive deeper into the influence of SML on DNN in section \ref{rq3}. 

Finally, we got a new state of the art that is DCN-V2 with SML achieved the best performance with AUC increased by about 0.03\% to 0.8117.

\begin{table}[]
\caption{Ablation Study}
\begin{tabular}{l l c c}
\hline
\toprule
&method       & Criteo                                        &  Avazu                        \\ \midrule
    Skip &DNN & 0.80948  & 0.78376            \\  
    & Vanilla & {\bf 0.80985} & {\bf 0.78409}
         \\ \hline
 &Relu & 0.80963  &  0.78357           \\   
     Activation            &       Sigmoid       &    0.80952        & 0.78390       \\
                 &        Tanh         &  {\bf 0.80978}   & {\bf 0.78407}                           \\ \hline
 & Weight(tanh)        & 0.80987            & 0.78403            \\  
 & Meta vanilla        & 0.81024            & 0.78426            \\ 
Scale & Meta Relu       & 0.80998          & 0.78384            \\\
           &   Meta Sigmoid     &  0.80984            & 0.78391            \\ 
 & Meta Tanh       & {\bf 0.81042}          & \textbf{0.78459}            \\  \bottomrule
\end{tabular}
\\
\vspace{1ex}
{The table gives ablation study of different modifications, which experimentally proofed the validity of proposed method.}
\label{Ablation}
\end{table}

\subsection{ Ablation Study(RQ2)}\label{rq2}
The influence of different improvements will be given in this part. The ablation study is based on DNN to exclude extra influence, we also defined a vanilla network whose  $skiplogit = W\odot x$ for clarification, which means each skip logit is directly weighted sum to final logit. Then activation function and meta scale are added to benefit the training of the model. 
\subsubsection{benefit of skip logit}
First of all, we perform an ablation study to identify the improvement brought by pure skip logit architecture. From Table \ref{Ablation}, simply sum and adding each layer's embedding to the final logit provides a moderate improvement on the Criteo dataset and Avazu dataset. To our surprise, directly adding Relu and Sigmoid activation to the vanilla will harm the performance, this may be due to the change of distribution that makes the identity hard. 
\subsubsection{benefit of Tanh}
Then we compared the performance under different activation functions. Relu, Tanh, and Sigmoid are chosen and here the sigmoid is to identify the benefit brought by standardization. The activation function is compared both under the condition that with and without meta scaling, the result is recorded in Table \ref{Ablation}. When activation function is added to input, the function will be ($f(x) = W\odot act(x)$). On Criteo dataset, Tanh slightly outperformed Relu and Sigmoid. A similar trend happens simultaneously on Avazu dataset. The largest improvement occurs when meta-scale information is introduced, the Tanh activation achieved the best performance with variance normalization alleviated the saturation. As a result, we verified the benefit introduced by Tanh. 
\subsubsection{Meta module Analysis}\label{meta_analysis}
We first compare the meta scale with simply learned parameters via backpropagation. In Table Table \ref{Ablation}, The meta Tanh outperforms weight Tanh by about 0.006\%, which can be considered as a great improvement. We also observed that the meta scale alleviates Tanh's saturation problem and achieved the best AUC 0.81042. Compared with the original DNN, the performance increased sharply by about 0.01\%. 

\subsection{Property Analysis(RQ3)}\label{rq3}
In this section, we explore how SML affects DNNs with neuron-level comparisons, and experimentally demonstrate the stability of SML. Without loss of generality, we set the number of hidden units to 400 in the following experiments. 
\begin{figure}[h]
\centering
\includegraphics[width=0.85\linewidth]{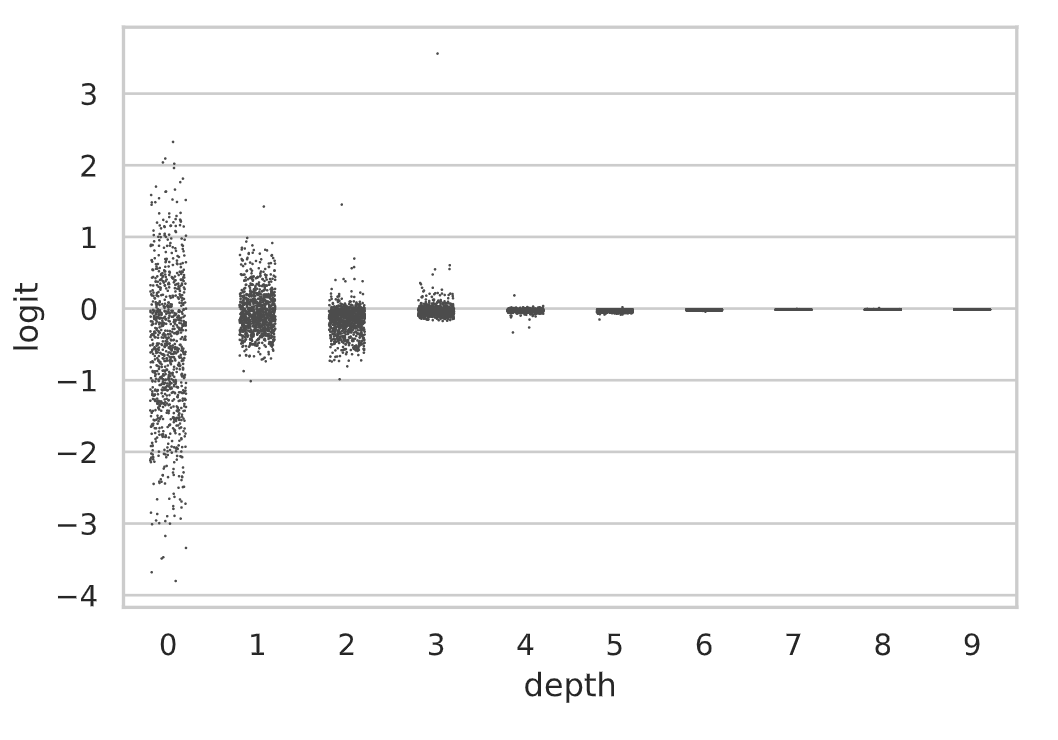}
\caption{Logit distribution in different depths
}
\label{log_var}
\end{figure}

\subsubsection{Logit Distribution.\label{logit_dist}} Figure \ref{log_var} shows the logit distribution of each layer. It can be seen that the variance of logit decreases as the network goes deeper. The variance of different samples almost fades away when the network reaches a depth of 5. 
This indicates shallow layers play important roles in the final prediction\cite{ensemble}. Skip-logit allows information to flow directly from shallow layers to final output, which emphasizes this nature and eases the training\cite{resnet_v2}.
In other extensive ctr prediction studies, the maximum number of DNN depth hardly goes beyond 5, which also supports our finding that shallow networks dominate the ctr prediction model.

\subsubsection{Dead Neurons.}  
\begin{figure}[h]
\centering
\includegraphics[width=\linewidth]{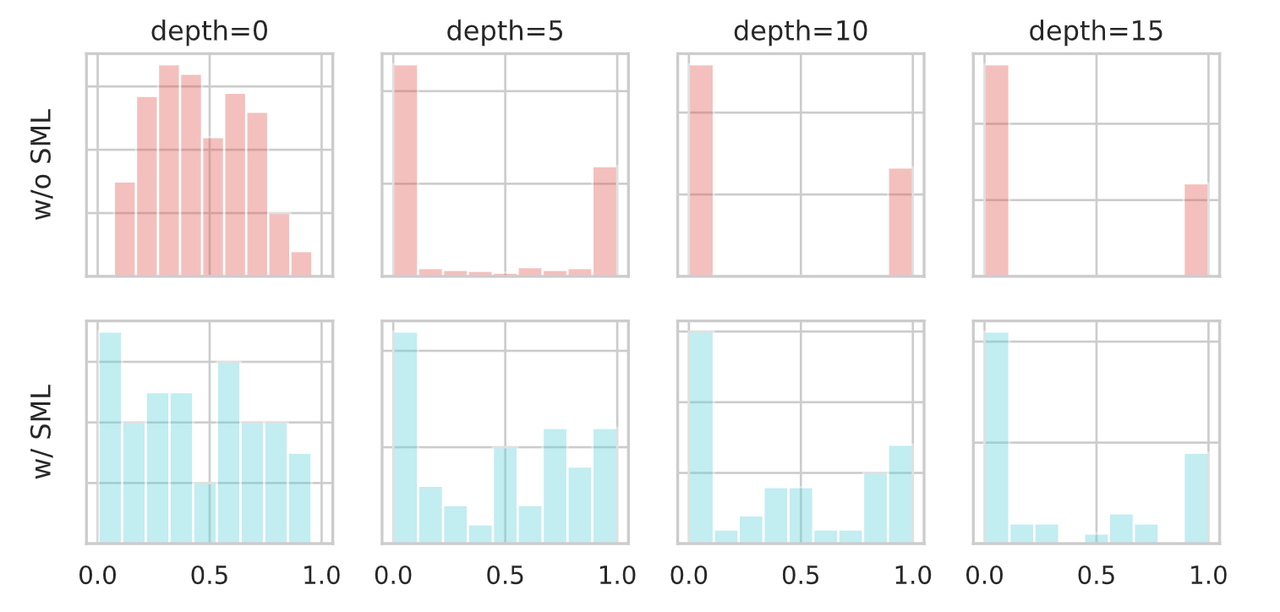}
\caption{Activation of rectifiers in deep networks. Upper: without SML, the kernel quickly become a bipolar distribution. Lower: with SML, a deeper layer's kernel keeps alive and provide information.
}

\label{kernel_dead}
\end{figure}

We investigate the activation status of neurons of DNNs w/ and w/o SML at different depths, and the results are demonstrated in Figure \ref{kernel_dead}.  In both settings, the pattern of activation status is shifting to a bipolar distribution with the growth of depth. This means that the state of the neuron is fixed, either always activated or dead, causing the degradation of the network's expressive ability\cite{sharp-gradient}. This further supports our previous observation that the effect of the deep final output is minimal.
However, if we take a closer look, it emerges that the distribution has already fallen into bipolarity at the 5-th layer for DNNs w/o SML, while staying flexible to some extent at 10-th layer of DNNs w/ SML. 
In addition to bridging between shallow layers and the final output, SML also has the effect of keeping the network alive at deeper layers, thereby gaining expressiveness. 

\begin{figure}[h]
\centering
\includegraphics[width=0.85\linewidth]{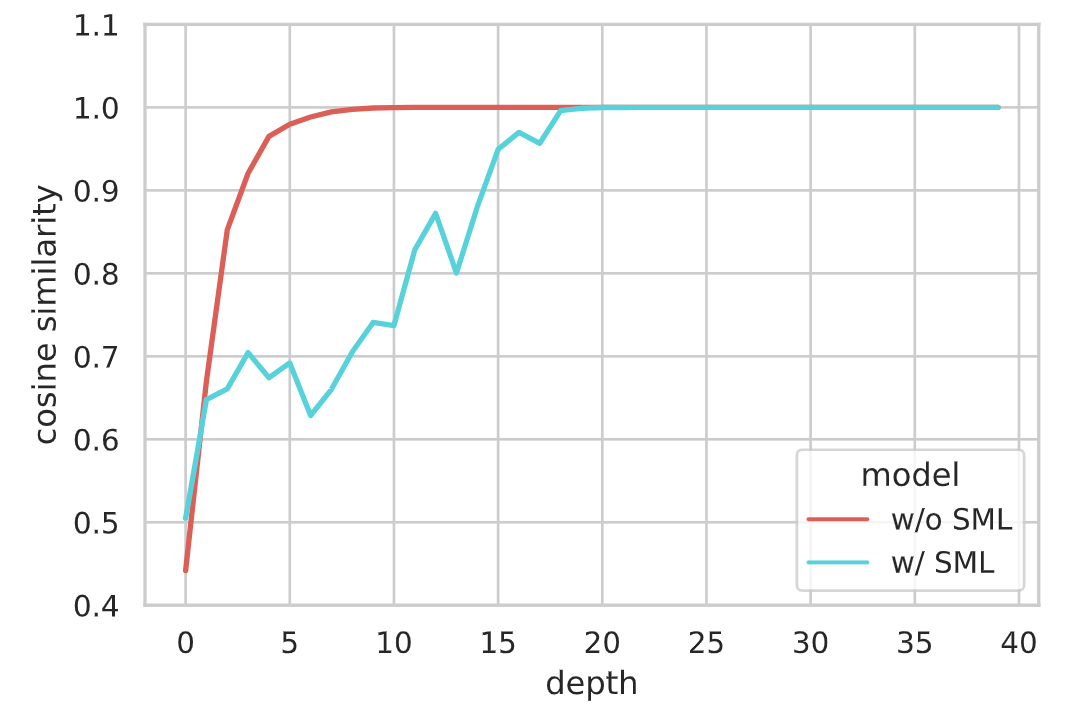}
\caption{Cosine similarity between different samples. Without SML, different input samples become indistinguishably similar in deeper layers. SML helps reflect information about the input data. 
}
\label{cos_sim}
\end{figure}

\subsubsection{Representation Similarity.} To directly reveal what the network learns, we calculate the average cosine similarity between different samples at different depths. The result is illustrated in Figure \ref{cos_sim}. If the similarity is equal to 1, it means that there is no distinction between different samples. As we can see, the similarity of DNN without SML is almost 1 at depth 5, which means that continuing to stack more layers can no longer improve the network's ability to distinguish samples. This issue is also addressed by \cite{collapse_bn} known as rank collapse, where activations for different input samples become indistinguishably similar in deeper layers. This can significantly slow training as the gradient updates no longer reflect information about the input data. SML effectively mitigates this problem, and the network does not crash until the 15th layer. The conclusion is cross-corroborated by the previous point.

\begin{figure}[h]
\begin{minipage}[t]{0.5\linewidth}
\centering
\setlength{\belowcaptionskip}{-0.1cm}   
\captionsetup{font={footnotesize}} 
\includegraphics[height=2.8cm,width=4.5cm]{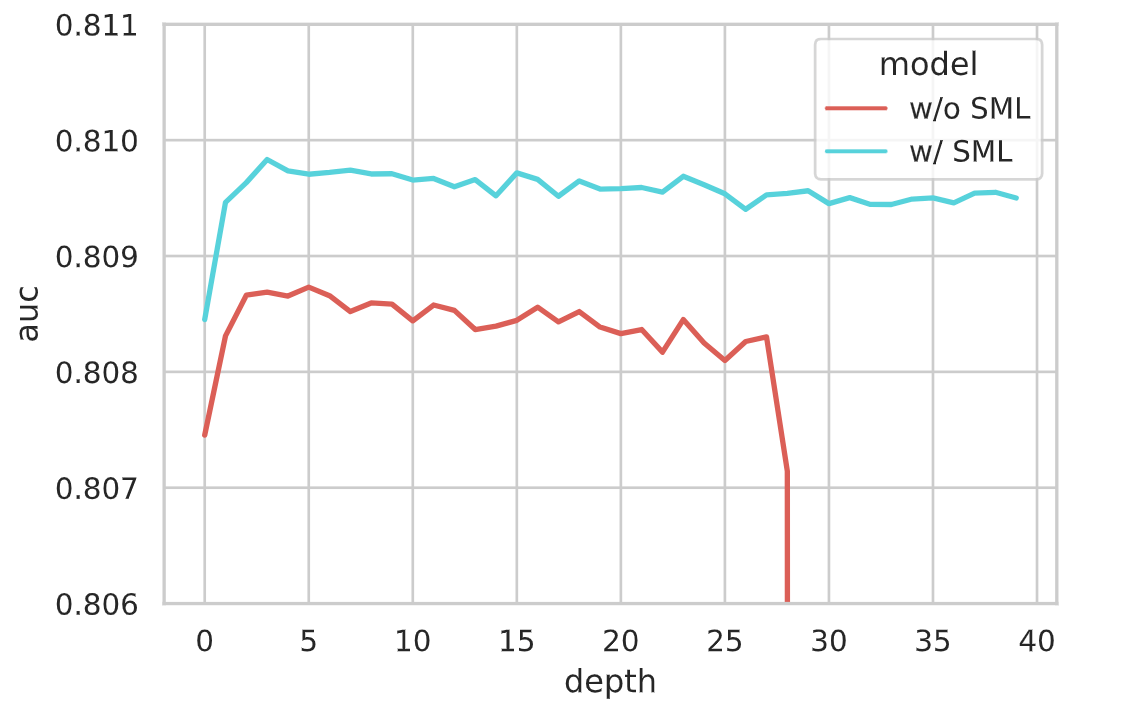}
\end{minipage}%
\begin{minipage}[t]{0.5\linewidth}
\captionsetup{font={footnotesize}}
\centering
\includegraphics[height=2.8cm,width=4.5cm]{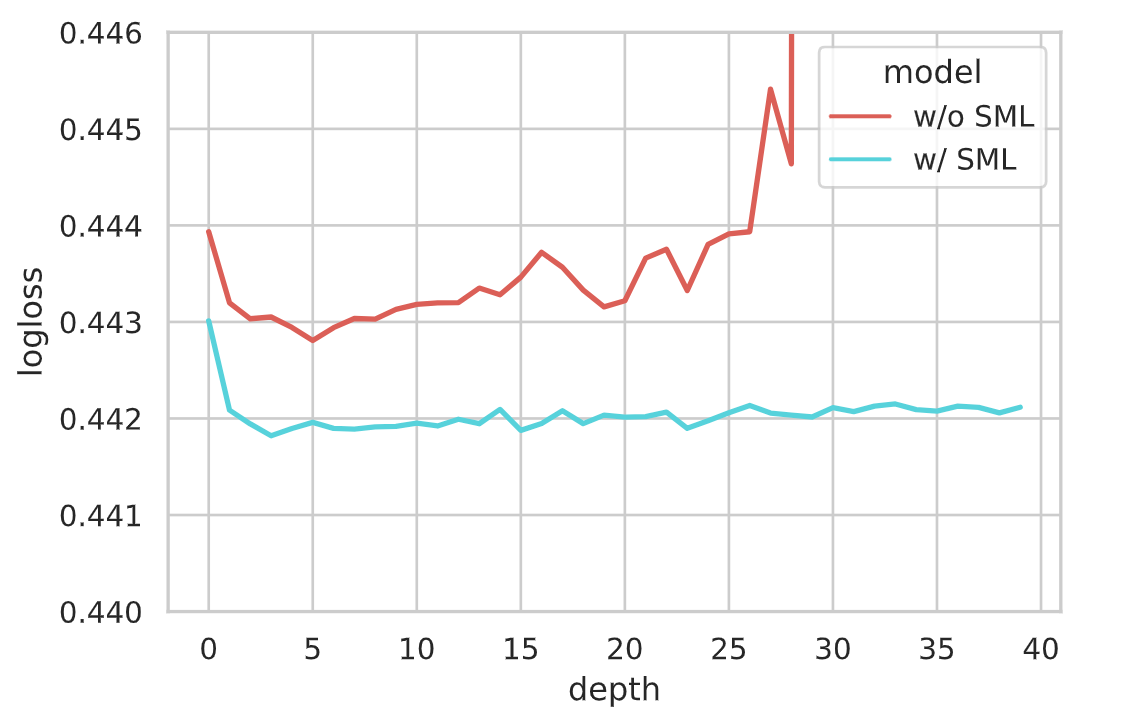}
\end{minipage}
\caption{ AUC and Logloss under different depthes on Criteo dataset. The proposed Skip Logit network can achieve stable performance under deep situation. However, DNN failed in training when the depth exceed 27.
}
\label{depth_training}
\end{figure}

\subsubsection{Training Stability.} We finally examined the convergence of DNN w/ SML and w/o SML at different depths, from 1 to 100 layers specifically. AUC and Logloss at each depth are illustrated in Figure \ref{depth_training}. Note that data points after depth 40 are omitted because the curve remains almost unchanged.

There are three phenomenons worth noting. First, SML undoubtedly boosts performance of DNN and both reach the best metric at depth of 4. Secondly, continuing increase the depth after 4 layers will result in diminishing effect, however milder symptoms of SML. Third, the training process of vanilla DNNs completely collapses at depth 27, while SML can stably train networks up to depth 100. The above three points all show the superb stability and convergence of SML.

\subsection{Online A/B Test(RQ4)}

We deploy this model to two different industrial scenarios at XXXX's online advertising system.
One is a vertical scenario for XXXX, and another is the main traffic of all ads. 
We run seven days' A/B Tests for each.
The traffic in both scenarios is equally divided into 50/50 for treatment and control groups.
In practice, we use AUC as the offline model metric and advertiser value\footnote{In general, ADVV is the bidding price given by the advertiser multiplied by the number of conversions. This is considered to be a metric that measures the total value that advertisers achieve from ads. Even though the specific definition of bidding and conversion may change according to pricing type, the meaning of ADVV stays still.}(ADVV) as the online business metric.
The improvement of the treatment group relative to the control group is listed in \ref{abtest}.
The uplift of both metrics proves the business value of SML in the real world.

\begin{table}[h]
\caption{Online A/B Test}
\begin{tabular}{l c c}
\hline
\toprule
Scenario &AUC &ADVV                       
\\ \midrule
XXXX & +0.07\% &+3.28\%
\\
Main Traffic &+0.15\%  &+1.13\%
\\  \bottomrule
\end{tabular}
\label{abtest}
\end{table}

\section{Conclusion}
In this paper, we proposed a SML module to introduce skip connection mechanism into the CTR prediction model thus improving DNN's optimization ability and fully stimulating DNN's expressive ability. With the help of theoretical proof and experimental analysis, the benefit of the proposed module is supported. By incrementally adding the SML module to the selected state of the art methods, consistent performance gains are observed.  For future work, we are interested in advancing our understanding of 1) how the weights of SML neural networks change; 2). introduce online variance statistical information to the meta normalization module.

\bibliography{sample-base}


\clearpage
\appendix

\section{Appendices}
\subsection{Proof of output variance}\label{variance}
First of all, we assume the input $x$ is derived from a Gaussian distribution $\mathcal{N}(0,\delta)$ .
\begin{align*}
E(relu(x)) =& \int_{-\infty}^{\infty}max(0,x)\frac{1}{\sqrt{2\pi}\delta^2}e^{\frac{-x^2}{2\delta^2}}dx \\
&=\int_{0}^{\infty}x\frac{1}{\sqrt{2\pi}\delta}e^{-\frac{x^{2}}{2\delta^{2}}}dx \\
&=\frac{1}{2}\int_{0}^{\infty}\frac{1}{\sqrt{2\pi}\delta}e^{-\frac{x^{2}}{2\delta^{2}}}dx^2 \\ 
&= \frac{\sqrt{2}\delta}{\sqrt{\pi}}\int_{0}^{\infty}e^{-\frac{x^{2}}{2\delta^{2}}}d\frac{x^{2}}{2\delta^{2}} \\
&= \frac{\sqrt{2}\delta}{\sqrt{\pi}} \\
Var(Relu(x)) &= E((x-\frac{\sqrt{2}\delta}{\sqrt{\pi}})^{2}) = E(x^{2}-2\frac{\sqrt{2}\delta}{\sqrt{\pi}}x + \frac{2\delta^{2}}{\pi}) \\
&\leq E(x^{2}) - \frac{2\delta}{\pi} = \delta^{2}(1-\frac{2}{\pi})
\end{align*}

\subsection{Proof of Theorem 1}\label{theorem1}
In the beginning, we start off with a simple claim that simplifies the population risk. We use $⟨A, B⟩$ denotes the inner product of A and B in the standard basis (that is, $⟨A, B⟩ = tr(A^{T}B)$ where $tr$ denotes the trace of a matrix.)
\\
{\bf Claim 1.} \textit{In the setting of this section, we have,}
\begin{equation}
f(A) = ||((I+A_{1}(I+A_{2}(I+...(I+A_{l}))))-R)\Sigma^{1/2}||^{2}_{F} + C
\end{equation}
\textit{Here $C$ is a constant that doesn’t depend on $A$, and $\Sigma^{1/2}$ denote the square root of ${\sigma}$, that is, the unique symmetric matrix $B$ that satisfies $B^{2} = \Sigma$.}
\\
\textit{Proof of Claim 2.} Let $tr(A)$ denotes the trace of the matrix $A$. Let $E = (I+A_{1}(I+A_{2}(I+...(I+A_{l}))))-R$. Recalling the definition of f (A) and using the equation \ref{optimize}, 
\begin{equation}\label{optimize}
f(A,(x,y)) = ||\hat{y} - y||^{2} =  ||((I+A_{1}(I+A_{2}(I+...(I+A_{l}))))x-Rx - \xi||
\end{equation}
we have
\begin{align*}
f(A)&= \mathbb{E}[||Ex − \xi||^{2}]\\
    &= \mathbb{E}[||Ex||^{2} + ||\xi||^{2}-2<Ex,\xi>]\\
    &= \mathbb{E}[||Ex||^{2} + ||\xi||^{2}-2\mathbb{E}<Ex,\mathbb{E}[\xi|x]>] \\ 
    &=E[tr(Exx^{T}E^{T})] + \mathbb{E}[||\xi||^{2}]\\
    &=tr(E\mathbb{E}[xx^{T}]E^{T})+C \\
    &=tr(E\Sigma E^{T})+C = ||E\Sigma^{1/2}||^{2}_{F}+C
\end{align*}
Then the gradient of object function $f$ is calculated. Let $ \bigtriangleup_{j} \in R^{d\times d}$ be a small change to the parameter $A_{j}$. With the help of Claim 1,
\begin{align*}
&f(A1,...,A_{j}+\bigtriangleup_{j},...,A_{l})\\
&= ||((I+A_{1}(...(I+A_{j}+\bigtriangleup_{j}(...(I+A_{l}))))-R)\Sigma^{1/2}||^{2}_{F}\\
&=||((I+A_{1}(...(I+A_{l})))-R)\Sigma^{1/2}+\bigtriangleup_{j}(I+A_{j+1}(...(I+A_{l})))\Sigma^{1/2}||^{2}_{F}\\
&=||((I+A_{1}(...(I+A_{l})))-R)\Sigma^{1/2}||^{2}_{F}+\\
&\quad 2<((I+A_{1}(...(I+A_{l})))-R)\Sigma^{1/2},\bigtriangleup_{j}(I+A_{j+1}(...(I+A_{l})))\Sigma^{1/2}>\\
&\quad +O(||\bigtriangleup_{j}||^{2}_{F})\\
&=f(A) + 2<E\Sigma(I+A_{j+1}(...(I+A_{l}))),\bigtriangleup_{j}>+O(||\bigtriangleup_{j}||^{2}_{F})
\end{align*}
By definition, this means that the $\frac{\partial f}{\partial A_{j}}=2<E\Sigma(I+A_{j+1}(...(I+A_{l}))),\bigtriangleup_{j}>$.
Then we can proof the global optima,
\begin{align*}
||\frac{\partial f}{\partial A_{i}}||_{F}&= 2||<E\Sigma(I+A_{j+1}(...(I+A_{l}))),\bigtriangleup_{j}>||_{F}\\
&\geq 2\delta_{min}(I+A_{i\backsim L}^{T})\delta_{min}(\Sigma)||E||_{F} \\
&\geq 2(1-\gamma_{i})\delta_{min}(\Sigma^{1/2})||E\Sigma^{1/2}||_{F},(\delta_{min}(I+A)\geq 1-||A||)
\end{align*}
For $||\bigtriangledown f(A)||^{2}_{F}$,
\begin{align*}
||\bigtriangledown f(A)||^{2}_{F} &=\sum^{l}_{i}||\frac{\partial f}{\partial A_{i}}||_{F}^{2}\\
&\geq 4\sum^{l}_{i}(1-\gamma_{i})\delta_{min}(\Sigma)||E\Sigma^{1/2}||_{F}^{2}\\ 
&\geq 4\sum^{l}_{i=1}(1-\gamma_{i})^{2} \delta_{min}(\Sigma)||f(A) - C||^{2}_{F}\\
&\geq 4\sum^{l}_{i=1}(1-\gamma_{i})^{2} \delta_{min}(\Sigma)||f(A) - C_{opt}||^{2}_{F}\\
&\quad (C_{opt}:=min_{A}f(A)\geq C)
\end{align*} 
Therefore we complete the proof of Theorem 1. To be clear, we recursively use the $I + A$ and the $\gamma^{i}$ is from the residual part $R^{i}$ that generated at each stage, where the $R^{i} = (R^{i-1}-I)A_{i-1}^{-1}$.  

\end{document}